\documentclass[12pt]{article}
\usepackage{amsmath}
\usepackage{psfrag}
\usepackage{graphicx}
\usepackage{graphics}
\usepackage{epsfig}
\newcommand{\ps}{p\!\!\!/}

\date{}
\begin{document}
\baselineskip=18.6pt plus 0.2pt minus 0.1pt \makeatletter


\title{\vspace{-3cm}
\hfill\parbox{4cm}{\normalsize \emph{}}\\
 \vspace{1cm}
{ Semirelativistic $1s-2s$ excitation of atomic hydrogen by electron impact }}
 \vspace{2cm}
\author{S. Taj$^1$, B. Manaut$^{1}$\thanks{{\tt b.manaut@usms.ma}} and L. Oufni$^2$\\
{\it {\small $^1$ Universit\'e Sultan Moulay Slimane, FPBM, LIRST,  BP : 523, 23000,  B\'eni Mellal, Morocco. }}\\
 {\it {\small $^2$ Universit\'e Sultan Moulay Slimane, FSTBM, LPMM,  BP : 523, 23000,  B\'eni Mellal, Morocco. }}}
   \maketitle \setcounter{page}{1}
\begin{abstract}
In the framework of the first Born approximation, we present   a
semirelativistic theoretical study of the inelastic excitation
($1s_{1/2}\longrightarrow 2s_{1/2}$) of hydrogen atom by electronic
impact. The incident and scattered electrons are described by a free
Dirac spinor and the hydrogen atom target is described by the Darwin
wave function.  Relativistic and spin effects are examined in the
relativistic regime. A detailed study has been devoted to the
nonrelativistic regime as well as the moderate relativistic regime.
Some aspects of this dependence as well as the dynamic behavior of
the DCS in the relativistic regime have been addressed.\\
PACS number(s): 34.80.Dp, 12.20.Ds
\end{abstract}
\section{Introduction}
The theoretical study of relativistic electron-atom collisions is
fundamental to our understanding of many aspects in plasma physics
and astrophysics. The development of electron-atom collision studies
has also been strongly motivated by the need of data  for testing
and developing suitable theories of the scattering and collision
process, and providing a tool for obtaining detailed information on
the structure of the target atoms and molecules. Many authors have
studied this process using numerical tools. Thus, Kisielius
\textit{et al.} \cite{1} employed, the R-matrix method with
nonrelativistic and relativistic approximations for the hydrogen
like $He^+$, $Fe^{25+}$ and $U^{91+}$ ions, where the case of
transitions $1s\longrightarrow 2s$ and $1s\longrightarrow 2p$ as
well as those between fine structure $n=2$ levels was considered.
Andersen \textit{et al.} \cite{2} have applied the semirelativistic
Breit Pauli R-matrix to calculate the electron-impact excitation of
the $^2S_{1/2}$ $\longrightarrow$ $^2P^o_{1/2,3/2}$ resonance
transitions in heavy alkali atoms. Payne \textit{et al.} \cite{3}
have studied the electron-impact excitation of the
$5s\longrightarrow 5p$ resonance transition in rubidium by using a
semi-relativistic Breit Pauli R-matrix with pseudo-states
(close-coupling) approach. Attaourti \textit{et al.} \cite{4} have
investigated the exact analytical relativistic excitation
$1S_{1/2}\longrightarrow 1S_{1/2}$ of atomic hydrogen, by electron
impact in the presence of a laser field. They have found that a
simple formal analogy links the analytical expressions of the
unpolarized differential cross section without laser and the
unpolarized differential cross section in the presence of a laser
field.

The aim of this contribution is to add some new physical insights
and to show that the non-relativistic formalism becomes enable to
describe particles with hight kinetic energies. Before we present
the results of our investigation, we first begin by sketching the
main steps of our treatment. For pedagogical purposes, we begin by
the most basic results of our work using atomic units (a.u) in which
one has ($\hbar=m_e=e=1$), where $m_e$ is the electron  mass at
rest, and which will be used throughout this work. We will also work
with the metric tensor $g^{\mu\nu}= diag(1,-1,-1,-1)$ and the
Lorentz scalar  product which is defined by $(a.b)=a^\mu b_\mu$. The
layout of this paper is as follows. We present the necessary
formalism of this work in section [2,3 and 4], the result and
discussion in section 5 and we end by a brief conclusion in section
6.
\section{Theory of the inelastic collision $1s_{1/2}\longrightarrow 2s_{1/2} $}
In this section, we calculate the exact analytical expression of the
semirelativistic unpolarized DCS for the relativistic excitation of
atomic hydrogen by electron impact. The transition matrix element
for the direct channel (exchange effects are neglected) is given by
\begin{eqnarray}
S_{fi} &=&-i\int dt\langle\psi _{p_{f}}(x_{1})\phi _{f}(x_{2})\mid
V_{d}\mid \psi _{p_{i}}(x_{1})\phi _{i}(x_{2})\rangle
 \nonumber \\
 &=&-i\int_{-\infty }^{+\infty }dt\int d\mathbf{r}_{1}\overline{\psi }
_{p_{f}}(t,\mathbf{r}_{1})\gamma^{0}\psi _{p_{i}}(t,\mathbf{r}_{1})
\langle\phi _{f}(x_{2})\mid V_{d}\mid \phi _{i}(x_{2})\rangle
\label{1}
\end{eqnarray}
where
\begin{equation}
V_{d}=\frac{1}{r_{12}}-\frac{Z}{r_{1}}  \label{2}
\end{equation}
is the direct interaction potential, $\mathbf{r}_{1}$ are the
coordinates of the incident and scattered electron, $\mathbf{r}_{2}$
 the atomic electron coordinates, $r_{12}=$ $\mid
\mathbf{r}_{1}-\mathbf{r}_{2}\mid $ and $r_{1}=\mid
\mathbf{r}_{1}\mid $. The function $\psi _{p_{i}}(x_{1})=\psi
_{p}(t,\mathbf{r}_{1})=u(p,s)\exp (-ip.x)/\sqrt{2EV}$ is the
electron wave function, described by a free Dirac spinor normalized
to the volume $V$, and $\phi _{i,f}(x_{2})=\phi
_{i,f}(t,\mathbf{r}_{2})$ are the semirelativistic wave functions of
the hydrogen atom where the index $i$ and $f$ stand for the initial
and final states respectively. The semirelativistic wave function of
the atomic hydrogen is the Darwin wave function for bound states
\cite{5}, which is given by :
\begin{equation}
\phi _{i}(t,\mathbf{r}_{2})=\exp (-i\mathcal{E} _{b}(1s_{1/2})t)\varphi_{1s} ^{(\pm )}(%
\mathbf{r}_{2})  \label{3}
\end{equation}
where $\mathcal{E} _{b}(1s_{1/2})$ is the binding energy of the ground
state of atomic hydrogen and $\varphi_{1s} ^{(\pm
)}(\mathbf{r}_{2})$ is  given by :
\begin{equation}
\varphi_{1s} ^{(\pm
)}(\mathbf{r}_{2})=(\mathsf{1}_{4}-\frac{i}{2c}\mathbf{\alpha
.\nabla }_{(2)})u^{(\pm )}\varphi _{0}(\mathbf{r}_{2})  \label{4}
\end{equation}
it represents a quasi relativistic bound state wave function,
accurate to first order in $Z/c$ in the relativistic corrections
(and normalized to the same order), with $\varphi _{0}$ being the
non-relativistic bound state hydrogenic function. The spinors
$u^{(\pm )}$ are such that $u^{(+)}=(1,0,0,0)^{T}$ and
$u^{(-)}=(0,1,0,0)^{T}$ and represent the basic four-component
spinors for a particle at rest with spin-up and spin-down,
respectively. The matrix differential operator $\alpha.\Delta$ is
given by :
\begin{eqnarray}
\alpha.\Delta=\begin{pmatrix}
  0 & 0 & \partial_z & \partial_x-i\partial_y \\
  0 & 0 & \partial_x+i\partial_y & -\partial_z \\
  \partial_z & \partial_x-i\partial_y & 0 & 0 \\
 \partial_x+i\partial_y &- \partial_z & 0 & 0
\end{pmatrix}
\end{eqnarray}
For the spin up, we have :
\begin{eqnarray}
\varphi_{1s} ^{(+)}(\mathbf{r}_{2})&=&N_{D_1} \left(
\begin{array}{c}
 1 \\
  0 \\
  \frac{i}{2cr_2}z \\
  \frac{i}{2cr_2}(x+iy)
\end{array}\right)
\frac{1} {\sqrt{\pi }} e^{-r_{2}}
\end{eqnarray}
and for the spin down, we have :
\begin{eqnarray}
\varphi_{1s} ^{(-)}(\mathbf{r}_{2})&=& N_{D_1}\left(
\begin{array}{c}
 0 \\
  1 \\
  \frac{i}{2cr_2}(x-iy) \\
  -\frac{i}{2cr_2}z
\end{array}\right)
\frac{1} {\sqrt{\pi }} e^{-r_{2}}
\end{eqnarray}
where
\begin{equation}
N_{D_1}=2c/\sqrt{4c^{2}+1}  \label{7}
\end{equation}
is a normalization constant lower but very close to 1. Let us
mention that the function $\phi _{f}(t,\mathbf{r}_{2})$ in Eq. (1)
is the Darwin wave function for bound states \cite{6}, which is also
accurate to the order $Z/c$ in the relativistic corrections. This is
expressed as $\phi
_{f}(t,\mathbf{r}_{2})=\exp (-i\mathcal{E}_{b}(2s_{1/2})t)\varphi_{2s} ^{(\pm )}(%
\mathbf{r}_{2})$ with $\mathcal{E}_{b}(2s_{1/2})$ as the binding energy
of the $2s_{1/2}$ state of atomic hydrogen.
\begin{eqnarray}
\varphi_{2s} ^{(+)}(\mathbf{r}_{2})&=&N_{D_2} \left(
\begin{array}{c}
 2-r_2 \\
  0 \\
  \frac{i(4-r_2)}{4r_2c}z \\
  \frac{(4-r_2)}{4rc}(-y+ix)
\end{array}\right)
\frac{1} {4\sqrt{2\pi }} e^{-r_{2}}
\end{eqnarray}
 for the spin up and
\begin{eqnarray}
\varphi_{2s} ^{(-)}(\mathbf{r}_{2})&=& N_{D_2}\left(
\begin{array}{c}
 0 \\
  2-r_2 \\
  \frac{4-r_2}{4cr_2}(y+ix)\\
  i\frac{(r_2-4)}{4cr_2}z
\end{array}\right)
\frac{1} {4\sqrt{2\pi }} e^{-r_{2}}
\end{eqnarray}
for the spin down. The transition matrix element in Eq. (\ref{1})
becomes :
\begin{eqnarray}
S_{fi}&=&-i\int_{-\infty }^{+\infty }dt\int d\mathbf{r}_{1}
d\mathbf{r}_{2}\overline{\psi }
_{p_{f}}(t,\mathbf{r}_{1})\gamma^{0}\psi
_{p_{i}}(t,\mathbf{r}_{1}) \phi^{\dag} _{f}(t,r_{2})\phi
_{i}(t,r_{2}) V_{d} \label{10}
\end{eqnarray}
and it is straightforward to get, for the transition amplitude,
\begin{eqnarray}
S_{fi} &=&-i\frac{\overline{u}(p_f,s_f)\gamma^0
u(p_i,s_i)}{2V\sqrt{E_fE_i}}2\pi H_{inel}(\Delta)
\delta\big(E_f+\mathcal{E}(2s_{1/2})-E_i-\mathcal{E}(1s_{1/2})\big)\nonumber\\
\label{11}
\end{eqnarray}
where  $\Delta=|p_i-p_f|$ and $\gamma^0$ is the Dirac matrix.
Using the  standard technique of the QED, we find for the
unpolarized DCS
\begin{eqnarray}
\frac{d\overline{\sigma}}{d\Omega_f}&=&\frac{|\mathbf{p}_f|}{|\mathbf{p}_i|}\frac{1}{(4\pi
c^2)^2}\left(\frac{1}{2}\sum_{s_is_f}|\overline{u}(p_f,s_f)\gamma^0
u(p_i,s_i)|^2\right)\left|H_{inel}(\Delta)\right|^2 \label{12}
\end{eqnarray}
\section{Calculation of the integral part}
The function $H_{inel}(\Delta)$ is found if one performs the various
integrals :
\begin{eqnarray}
H_{inel}(\Delta)=\int_0^{+\infty}d\mathbf{r}_1 e^{i\mathbf{\Delta}
\mathbf{r}_1}I(\mathbf{r}_1)\label{13}
\end{eqnarray}
\subsection{Integral over $\mathbf{r}_2$}
The quantity $I(\mathbf{r}_1)$ is easily evaluated in the following
way. We first write the explicit form of $I(\mathbf{r}_1)$ :
\begin{eqnarray}
I(\mathbf{r}_1)=\int_0^{+\infty}d\mathbf{r}_2\phi^{\dag}
_{2s}(\mathbf{r}_{2})\left[\frac{1}{r_{12}}-\frac{Z}{r_1}\right]\phi _{1s}(\mathbf{r}_{2})
\label{14}
\end{eqnarray}
Next, we develop the quantity $r^{-1}_{12}$ in spherical harmonics
as
\begin{eqnarray}
\frac{1}{\mathbf{r}_{12}}=4\pi\sum_{lm}\frac{Y_{lm}(\widehat{r}_1)Y_{lm}^*(\widehat{r}_2)}{2l+1}\frac{(\mathbf{r}_<)^l}{(\mathbf{r}_>)^{l+1}}\label{15}
\end{eqnarray}
where $r_>$ is the greater of $r_1$ and $r_2$, and $r_<$ the lesser
of them. The angular coordinates of the vectors $\mathbf{r}_1$ and
$\mathbf{r}_2$ are such that : $\widehat{r}_1=(\theta_1,\varphi_1)$
and $\widehat{r}_2=(\theta_2,\varphi_2)$. We use the well known
integral \cite{7}
\begin{eqnarray}
\int_x^{+\infty} du \;u^{m} e^{-\alpha u}= \frac{m!}{\alpha^{m+1}} e^{-\alpha x}\sum_{\mu=0}^m \frac{\alpha^\mu x^\mu}{\mu ! }\qquad \qquad Re(\alpha)>0 \label{16}
\end{eqnarray}
then, after some analytic calculations, we get for $I(\mathbf{r}_1)$
:
\begin{eqnarray}
I(\mathbf{r}_1)=\frac{6}{27}(\frac{1}{c^2}-4)+\frac{4}{27c^2}\frac{1}{\mathbf{r}_1}-\frac{4}{9}(1+\frac{1}{8c^2})\mathbf{r}_1\label{17}
\end{eqnarray}
\subsection{Integral over $\mathbf{r}_1$}
The integration over $\mathbf{r}_1$ gives rise to the following formula :
\begin{eqnarray}
H_{inel}(\Delta)=\int_0^{+\infty}d\mathbf{r}_1 e^{i\mathbf{\Delta} \mathbf{r}_1} I(\mathbf{r}_1)=-\frac{4\pi}{\sqrt{2}}(I_1+I_2+I_3)
\label{18}
\end{eqnarray}
the angular integrals are performed by expanding the plane wave
$e^{i\mathbf{\Delta} \mathbf{r}_1}$ in spherical harmonics as :
\begin{eqnarray}
e^{i\mathbf{\Delta} \mathbf{r}_1} = \sum_{lm}4\pi i^l
j_l(\mathbf{\Delta}
\mathbf{r}_1)Y_{lm}(\widehat{\mathbf{\Delta}})Y_{lm}^*(\widehat{\mathbf{r}}_1)
\label{19}
\end{eqnarray}
with $\mathbf{\Delta}=\mathbf{p}_i-\mathbf{p}_f$ is the relativistic
momentum transfer and $\widehat{\mathbf{\Delta}}$ is the angular
coordinates of the vector $\mathbf{\Delta}$. Then, after some
analytic computations, we get for $I_1$, $I_2$ and $I_3$ the
following result :
{\small
\begin{eqnarray}
I_1&=&\frac{4}{27c^2}\int_0^{+\infty}dr_1\;r_1e^{-\frac{3}{2}r_1}j_0(\Delta r_1)=\frac{4}{27c^2}\frac{1}{((3/2)^2+\mathbf{\Delta}^2)}\nonumber\\
I_2&=&\frac{6}{27}(\frac{1}{c^2}-4)\int_0^{+\infty}dr_1\;r_1^2e^{-\frac{3}{2}r_1}j_0(\Delta r_1)=\frac{2}{27}(\frac{1}{c^2}-4)\frac{3}{((3/2)^2+\mathbf{\Delta}^2)^2}\\
\label{13}
I_3&=&-\frac{4}{9}(1+\frac{1}{8c^2})\int_0^{+\infty}dr_1\;r_1^3e^{-\frac{3}{2}r_1}j_0(\Delta
r_1)=\frac{8}{9}(1+\frac{1}{8c^2})\frac{\mathbf{\Delta}^2-27/4}{((3/2)^2+\mathbf{\Delta}^2)^3}\nonumber
\end{eqnarray}
}
It is clear that the situation is different than in elastic
collision \cite{4}, since we have no singularity in the case
($\mathbf{\Delta} \to 0$)
\begin{figure}[!b]
 \centering
\includegraphics[angle=0,width=3 in,height=3.5 in]{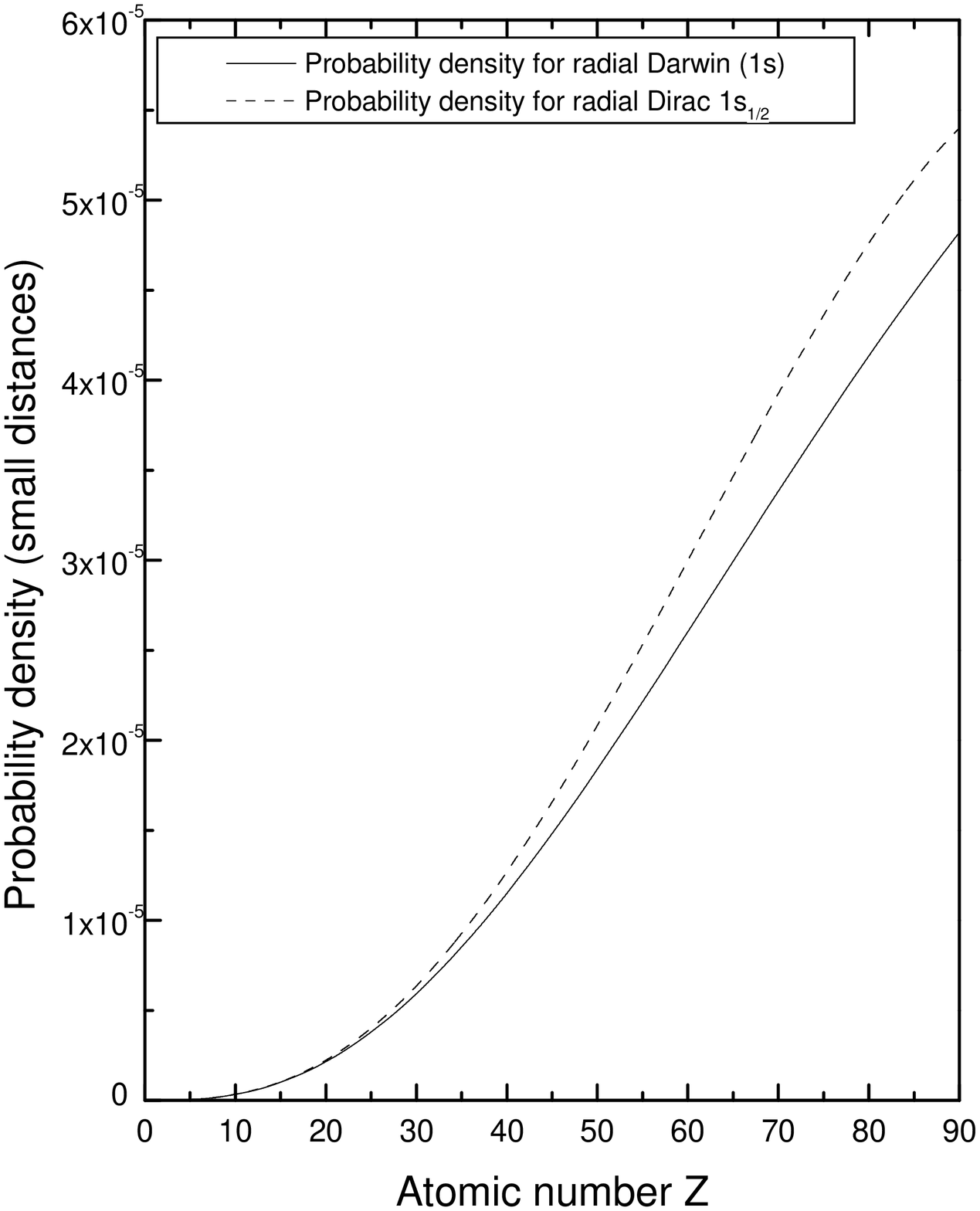}
\caption{\label{fig1}Behavior of the probability density for  radial
Darwin wave function compared with that of the Dirac wave function
for small distances and for increasing values of the atomic charge
number.}
\end{figure}
\section{Calculation of the spinorial part}
The calculation is now reduced to the computation of traces of
$\gamma$ matrices. This is routinely done using Reduce \cite{8}. We
consider the unpolarized DCS. Therefore, the various polarization
states
 have the same probability and the actual calculated spinorial part is given by summing over the final
  polarization $s_f$ and averaging aver the initial polariztion $s_i$. Therfore, the spinorial
  part is given by :
\begin{eqnarray}
\frac{1}{2}\sum_{s_is_f}|\overline{u}(p_f,s_f)\gamma^0
u(p_i,s_i)|^2&=&\text{Tr}\left\{\gamma^0(\ps_ic+c^2)\gamma^0(\ps_fc+c^2)\right\}\nonumber\\
&=&2c^2[\frac{2E_fE_i}{c^2}-(p_i.p_f)+c^2]\label{14}
\end{eqnarray}
We must, of course, recover the result in the nonrelativistic limit
($\gamma \longrightarrow 1$), situation of which the differential
cross section can simply given by :
\begin{eqnarray}
\frac{d\overline{\sigma}}{d\Omega_f}=\frac{|\mathbf{K}_f|}{|\mathbf{K}_i|}\frac{128}{\left(|\mathbf{\Delta_{nr}}|^2+\frac{9}{4}\right)^6}\label{14}
\end{eqnarray}
with $|\mathbf{\Delta_{nr}}|=|\mathbf{K}_i-\mathbf{K}_f|$ is the
nonrelativistic momentum transfer and the momentum vectors
($\mathbf{K}_i$, $\mathbf{K}_f$) are related by the following
formula :
\begin{eqnarray}
\mathbf{K}_f=(|\mathbf{K}_i|^2-3/4)^{1/2}
\end{eqnarray}
\begin{figure}[!b]
 \begin{minipage}[b]{.46\linewidth}
  \centering\epsfig{figure=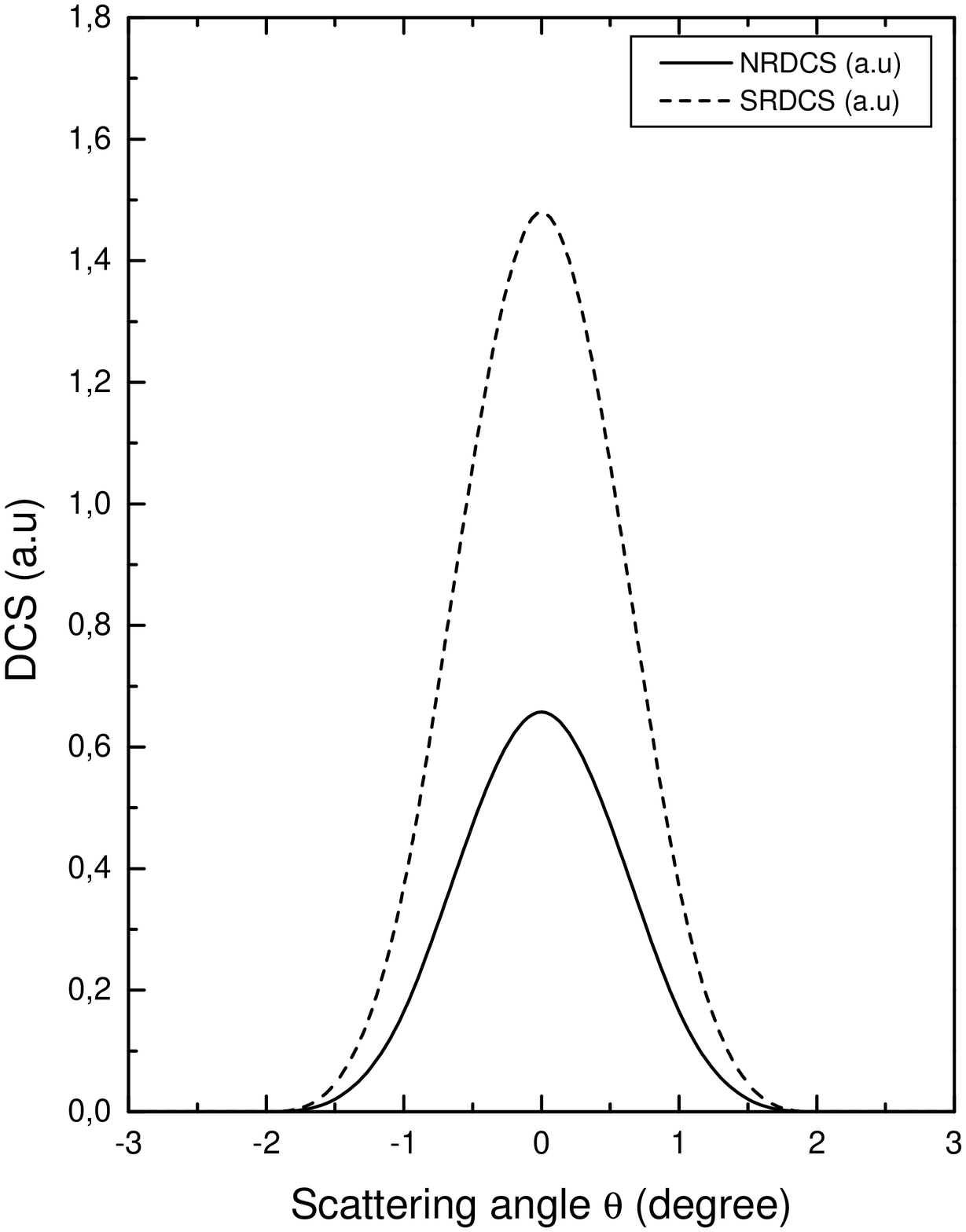,width=\linewidth,height=3.3in}
  \caption{\small{ \label{fig2} The long-dashed line represents the
semi-relativistic DCS, the  solid line represents the corresponding
non-relativistic DCS for a relativistic parameter ($\gamma=1.5$) as
functions of the scattering angle $\theta $. \vspace{0.4 cm} }}
 \end{minipage} \hfill
 \begin{minipage}[b]{.46\linewidth}
  \centering\epsfig{figure=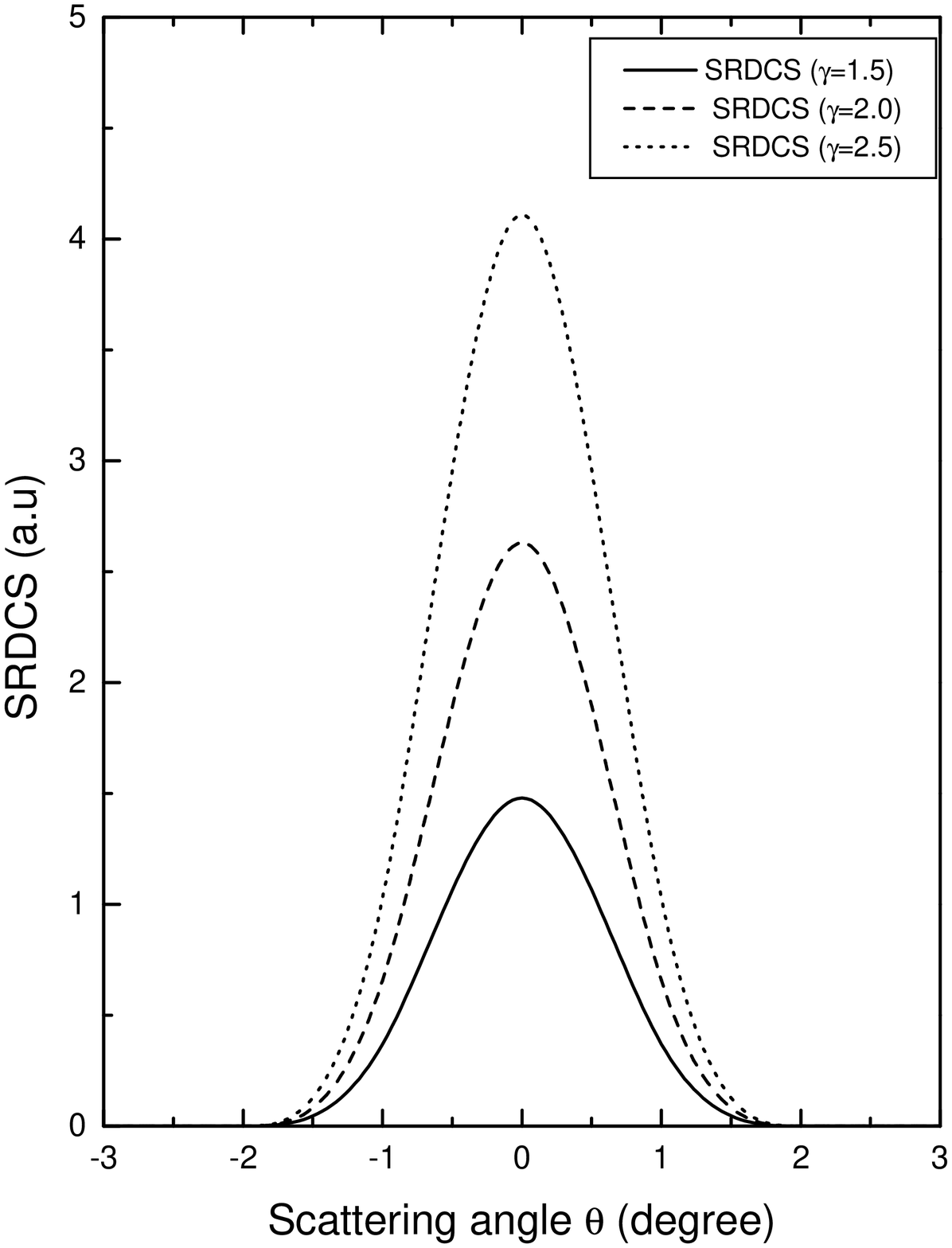,width=\linewidth,height=3.3in}
  \caption{\small{ \label{fig3} The solid line represents the
semi-relativistic DCS, the long-dashed line represents the
corresponding non-relativistic DCS  for various values of the
relativistic parameter ($\gamma=1.5$, $\gamma=2$ and $\gamma=2.5$)
as functions of the scattering angle $\theta$.}}
 \end{minipage}
\end{figure}
\section{Results and discussions}
In presenting our results it is convenient to consider separately
those corresponding to non-relativistic regime (the relativistic
parameter $\gamma \simeq 1$) and those related to relativistic one
(the relativistic parameter $\gamma \simeq 2$). Before beginning the
discussion of the obtained results, it is worthwhile to recall the
meaning of some abbreviation that will appear throughout this
section. The NRDCS stands for the nonrelativistic differential cross
section, where nonrelativistic plane wave are used to describe the
incident and scattered electrons. The SRDCS stands for the
semirelativistic differential cross section.\\
\indent We begin our numerical work, by the study of the dependence
of the probability density for radial Darwin and Dirac wave
functions, on the atomic charge number $Z$.
\begin{figure}[!b]
 \begin{minipage}[b]{.46\linewidth}
  \centering\epsfig{figure=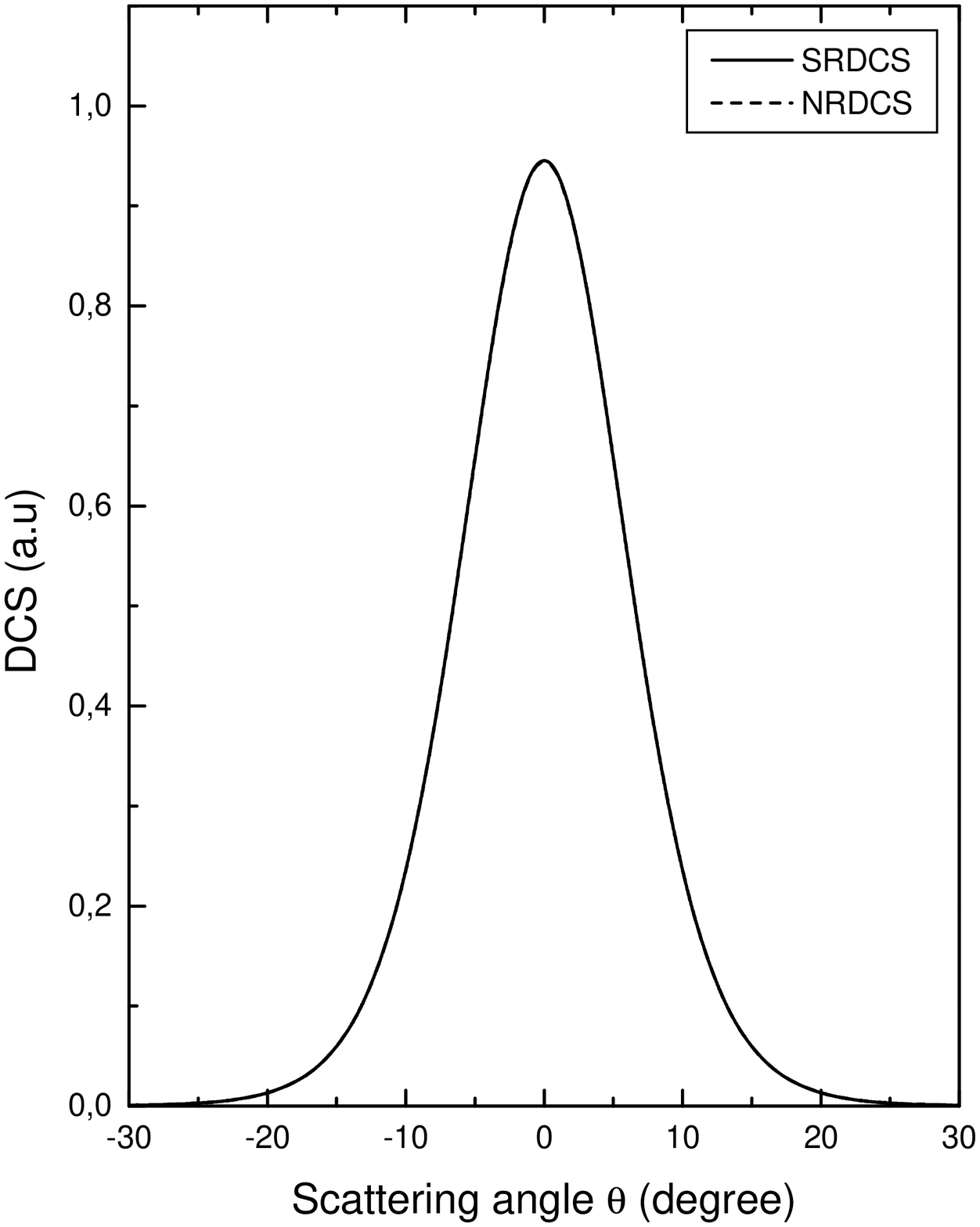,width=\linewidth,height=3.3in}
  \caption{\small{\label{fig4} The solid line represents the
semi-relativistic DCS, the long-dashed line represents the
corresponding non-relativistic DCS for a relativistic parameter
$\gamma=1.00053$ as functions of the scattering angle
$\theta$.\vspace{0.4cm}}}
 \end{minipage} \hfill
 \begin{minipage}[b]{.46\linewidth}
  \centering\epsfig{figure=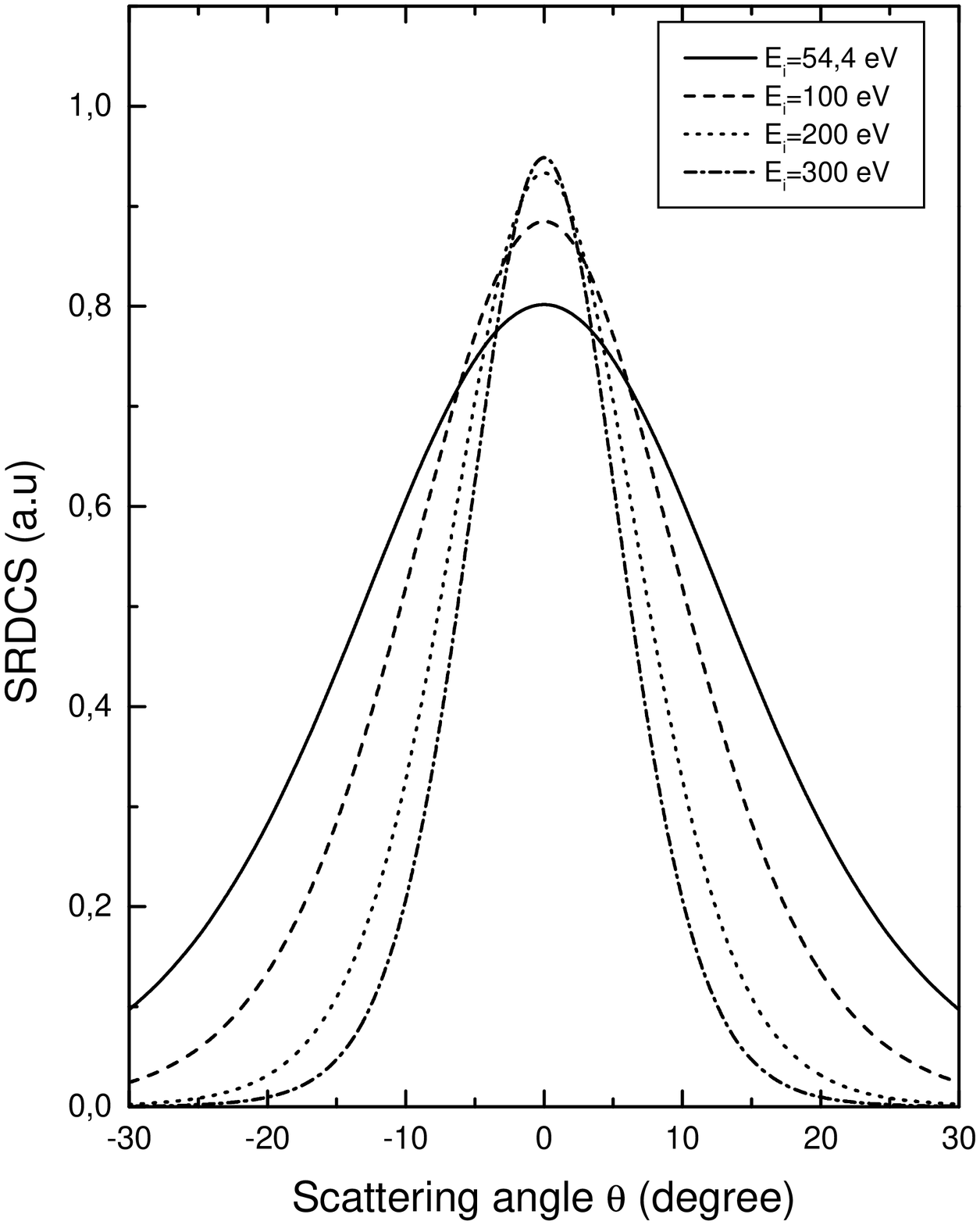,width=\linewidth,height=3.3in}
  \caption{\small{ \label{fig5} The variation of the SRDCSs with respect to $\theta$,
   for various kinetic energies.\vspace{1.8cm}}}
 \end{minipage}
 \end{figure}
As long as the condition $Z\alpha \ll 1$ is verified, the use of
Darwin wave function do not have any influence at all on the results
at least in the first order of perturbation theory. So, the
semi-relativistic treatment when $Z$ increases may generate large
errors but not in the case of this work. In this paper, we can not
have numerical instabilities since there are none. For the sake of
illustration, we give below the behavior of the probability density
for radial Darwin wave functions as well as that of the exact
relativistic Dirac wave functions for different values of $Z$. As
you may see, even if it is not noticeable on the figure 1, there are
growing discrepancies for $Z=10$ and these become more pronounced
when $Z=20$. The QED formulation shows that there are relativistic
and spin effects at the relativistic domain and the non relativistic
formulation is no
longer valid.\\
\begin{figure}[!t]
 \centering
\includegraphics[angle=0,width=3 in,height=3.3 in]{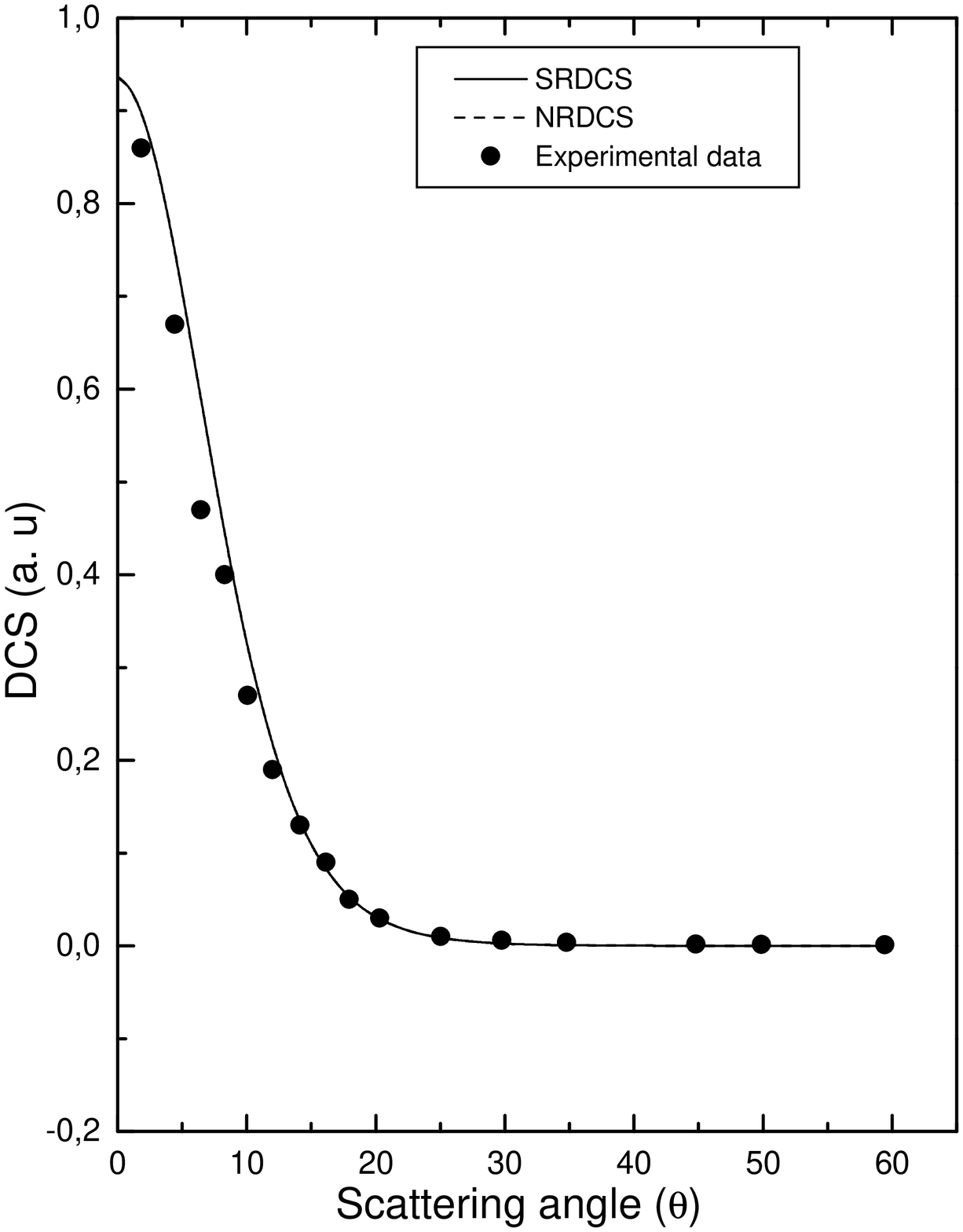}
\caption{\label{fig6} The variation of the differential $1s-2s$
cross section of $e^--H$ scattering at $200\; eV$. The dots are the
observed values of J. F. Wiliams (1981) ; the solid line represents
the semi-relativistic approximation and the long-dashed line
corresponds to the non-relativistic DCS.}
\end{figure}
\indent In the relativistic regime, the semirelativistic
differential cross section results obtained for the
$1s\longrightarrow 2s$ transition in atomic hydrogen by electron
impact, are displayed in figures 2 and 3. In this regime, there are
no theoretical models and experimental data for comparison as in
nonrelativistic regime. In such a situation, it appears from figures
\ref{fig2} and \ref{fig3} that in the limit of high electron kinetic
energy, the effects of the additional spin terms and the relativity
begin to be noticeable and that the non-relativistic formalism is no
longer applicable. Also a pick in the vicinity of
$\theta_f=0^{\circ}$ is clearly observed.\\
\indent The investigation in the nonrelativistic regime were
conducted with $\gamma$ as a relativistic parameter and $\theta$ as
a scattering angle. In atomic units, the kinetic energy is related
to $\gamma$ by the following relation : $E_k=c^2(\gamma-1)$. Figure
4 shows the dependence of DCS, obtained in two models (SRDCS,
NRDCS), on scattering angle $\theta$.  In this regime, it appears
clearly that there is no difference between these models. Figure 5
shows the variation of the SRDCS with $\theta$ for various energies.
It also shows approximatively in the interval [-5, 5], the SRDCS
increases with $\gamma$, but decreases elsewhere. Figure 6 presents
the observed and calculated angular dependence of $1s-2s$
differential cross section of $e^--H$ scattering at incident energie
$200\;eV$. Results obtained in two approaches semirelativistic and
non-relativistic approximations are indistinguishable and in good
agreement with the experimental data provided by J. F. Williams
\cite{9}.
\section{Conclusion}
In this paper we have presented the results of a semirelativistic
excitation of atomic hydrogen by electronic impact. We have used the
simple semirelativistic Darwin wave function that allows to obtain
analytical results in an exact and closed form within the framework
of the first Born approximation. This model gives good results if
the condition $Z/c\ll 1$ is fulfilled. We have compared our results
with previous nonrelativistic results and have found that the
agreement between the different theoretical approaches is good in
the nonrelativistic regime. We have also showed that the
non-relativistic treatment is no longer reliable for energies
higher. We hope that we will be able to compare our theoretical
results with forthcoming experimental data in the relativistic
regime.\\
\begin{center}
\textbf{ ACKNOWLEDGMENT}
\end{center}
I would like to thank Professor N. BOURIMA for his help in
maintaining the text linguistically acceptable.

\end{document}